\documentclass[aps,prb,twocolumn,superscriptaddress,showpacs,floatfix]{revtex4}
\usepackage{graphicx}

\newcommand{\be}{\begin{equation}}
\newcommand{\ee}{\end{equation}}
\newcommand{\bea}{\begin{eqnarray}}
\newcommand{\eea}{\end{eqnarray}}
\newcommand{\bi}{\bibitem}

\renewcommand{\r}{({\bf r})}
\newcommand{\rp}{({\bf r'})}

\newcommand{\epthree}{\epsilon_{TF}({\bf k})}
\newcommand{\eptwo}{\epsilon_{TF,2}({\bf k})}
\newcommand{\ttfthree}{T_{TF}[n]}
\newcommand{\ttftwo}{T_{TF,2}[n]}
\newcommand{\epdthree}{\epsilon_{TFD}({\bf k})}
\newcommand{\epdtwo}{\epsilon_{TFD,2}({\bf k})}

\newcommand{\epcthree}{\epsilon_{TFDc}({\bf k})}

\begin{document}

\title{Construction of model dielectric functions for two and three
dimensional electron liquids from density functionals}

\author{A. P. F\'avaro}
\affiliation{Departamento de F\'{\i}sica e Inform\'atica,
Instituto de F\'{\i}sica de S\~ao Carlos,
Universidade de S\~ao Paulo,
Caixa Postal 369, 13560-970 S\~ao Carlos, SP, Brazil}
\author{Jo\~ao V\'{\i}tor Batista Ferreira}
\affiliation{Departamento de F\'{\i}sica, CCET, Universidade Federal
de Mato Grosso do Sul, Campo Grande, MS, Brazil}
\author{K. Capelle}
\email{capelle@if.sc.usp.br}
\affiliation{Departamento de F\'{\i}sica e Inform\'atica,
Instituto de F\'{\i}sica de S\~ao Carlos,
Universidade de S\~ao Paulo,
Caixa Postal 369, 13560-970 S\~ao Carlos, SP, Brazil}
\date{\today}

\begin{abstract}
The Thomas-Fermi (TF) approximation for the static dielectric function of 
a three-dimensional (3d) electron liquid can be derived by minimizing the 
TF local-density approximation for the kinetic-energy functional. Here we 
show that this connection between energy functionals and model dielectric 
functions is not an artifact, but a general paradigm. Four examples are worked 
out in detail, by calculating the model dielectric functions that follow, 
respectively, from (i) exchange corrections to TF theory in 3d, i.e., TF-Dirac 
theory, (ii) further correlation corrections to TF-Dirac theory in 3d, (iii) 
TF theory in 2d, and (iv) exchange corrections to TF theory in 2d. Each of 
these cases has certain interesting features, revealing connections 
between independent many-body methods, that are discussed in some detail.
As a byproduct of these investigations we also find that a common textbook 
statement about the long-wavelength ($k\to0$) limit of the random-phase 
approximation is not fully correct.
\end{abstract}



\maketitle

\section{Introduction}
\label{intro}

Screening is one of the most important manifestations of many-body effects 
in metals, semiconductors and plasmas.\cite{pinesnozieres,quantliq}
In the Thomas-Fermi (TF) approximation,\cite{spruch} static screening in 
three-dimensional (3-d) electron liquids is described by means of a 
wave-vector-dependent dielectric function,\cite{ashcroft,ziman,phillips}
\be
\epthree=1+{k_{TF}^2\over k^2},
\label{eps3def}
\ee
where the TF screening wave vector is $k_{TF}=\sqrt{4 k_F/\pi a_0}$. Here
$k_F=(3 \pi^2 n_0)^{1/3}$ and $a_0=\hbar^2/m e^2$ denote the Fermi wave 
vector and the Bohr radius, respectively, and $n_0$ is the charge density 
of the unperturbed electron liquid.  This model dielectric
function is sometimes used in its own right as an approximation for
screening in metals, plasmas and doped semiconductors; more frequently, it
appears as starting point for, or limiting case of, more sophisticated
many-body treatments of screening.\cite{ashcroft,ziman,phillips,mahan,march}
We stress from the outset that our aim in the present work is not the 
construction of high-precision or material-specific expressions for the 
dielectric function, but to explore connections between different many-body
methods that become visible through the expressions they yield for the
dielectric functions.

The Thomas-Fermi approximation to the kinetic energy of a 3-d electron
liquid is \cite{dftbook,parryang}
\be
\ttfthree= \frac{3\hbar^2 (3\pi^2)^{2/3}}{10 m} \int d^3r\, n\r^{5/3}
= C \int d^3r\, n\r^{5/3}.
\label{tflda}
\ee

While both $\epthree$ and $\ttfthree$ are widely known and employed concepts, 
it is worthwhile stressing that the two uses of the label `Thomas-Fermi'
are not equivalent: the 3-d TF dielectric function can be obtained from the 
TF energy functional by one additional approximation, 
a linearization of the Euler equation resulting from minimizing 
$\ttfthree$.\cite{mahan,jonesmarch} To make our paper self-contained we start,
in Sec.~\ref{tfsec}, by briefly recalling this derivation. 

The inclusion of many-body effects beyond the TF approximation is in
general highly nontrivial, and typically accomplished by either constructing
local-field corrections to the random-phase approximation (RPA),\cite{mahan}
or by first-principles density-functional calculations of the dielectric
functions of realistic materials.\cite{abineps} In Sec.~\ref{tfdsec} we show 
that certain important many-body effects can be incorporated into the TF 
dielectric function in a much simpler way, by employing the Dirac-Slater 
local-density approximation (LDA) for the exchange 
energy,\cite{march,dftbook,parryang}
\bea
E_x^{LDA}[n]=  - \frac{3e^2}{4} \left(\frac{3}{\pi}\right)^{1/3} 
\int d^3r\, n\r^{4/3} 
\\
= - D \int d^3r\, n\r^{4/3}.
\label{dirac3d}
\eea
Addition of this to the TF energy functional leads to the so-called
Thomas-Fermi-Dirac (TFD) approximation.\cite{march,dftbook,parryang}
Further correlation corrections can be described by the local-density
approximation \cite{dftbook}
\be
E_{xc}^{LDA}[n]= E_x^{LDA}[n] + \int d^3r\, e_c(n)|_{n\to n\r},
\label{corrlda}
\ee
where $e_c(n)$ is the per-volume correlation energy of the uniform 
electron liquid \cite{ceperley} underlying common parametrizations
of the LDA.\cite{dftbook} In Sec.~\ref{tfdcsec} we derive the dielectric 
function arising from $E_{xc}^{LDA}$. Secs.~\ref{2dtfsec} and \ref{2dtfdsec}
extend these calculations to two-dimensional electron liquids, within TF 
and TFD theory, respectively. As a byproduct of this investigation we also
revisit, in Sec.~\ref{limit}, the long wavelength limit of the 
three-dimensional dielectric function found in the RPA, which has as 
nontrivial correction similar to the one found here within TFD theory.

\section{TF dielectric function in three-dimensional electron liquids}
\label{tfsec}

This derivation, although elementary, will be presented in some 
detail, because it serves as a model for the more complicated cases treated
afterwards. The 3-d TF total-energy functional is 
\bea
E_{TF}[n]=
\ttfthree+
\nonumber \\
\frac{e^2}{2}\int d^3r \int d^3r'
\frac{n\r n\rp}{|{\bf r}-{\bf r'}|} + \int d^3r \, n\r v_{ext}\r,
\label{etotal3}
\eea
where the second term on the right-hand-side (rhs) is the electrostatic
(Hartree) energy of the charge distribution $n\r$ with itself, and the
third term describes the potential energy of this charge distribution in
the external potential $v_{ext}\r$, composed of a smeared-out background of 
positive charge with constant density $n_{bg}$ and potential $v_{bg}$ 
and an additional test 
charge $n_{test}\r$ with potential $v_{test}\r$. It is the screening of this 
test charge, which is normally taken to be a point charge, that the theory 
aims to describe. Equilibrium is characterized by the condition
$\delta(E_{TF}[n]-\mu N)/\delta n\r = 0$,
where the total particle number $N=\int d^3r\, n\r$ and the chemical potential
$\mu$ at zero temperature is the Fermi energy $\epsilon_F$. This minimization
leads to the Euler equation
\be
\frac{5}{3}Cn\r^{2/3}+v_{ext}\r + v_H\r -\mu=0,
\label{euler}
\ee
where the Hartree potential is $v_H = \delta E_H[n]/\delta n = 
e^2\int d^3r'\, n\rp/|{\bf r}-{\bf r'}|$.
Solving Eq.~(\ref{euler}) for $n\r$ we find
\be
n\r=\left(\frac{3}{5C}\right)^{3/2}(\mu-v_s\r)^{3/2},
\label{eulern}
\ee
where we have introduced the effective, or screened, potential
$v_s=v_H+v_{ext}$. If this potential is weak, i.e., $v_s/\mu \ll 1$,
Eq.~(\ref{eulern}) can be linearized to give
\be
n\r=
\left(\frac{3\mu}{5C}\right)^{3/2}\left(1-\frac{3}{2}\frac{v_s\r}{\mu}\right).
\ee
By introducing the charge density of the unperturbed system,
$n_0=k_F^3/3 \pi^2$ (where $\hbar^2 k_F^2/2m = \mu$), and the corresponding
3-d density of states, $g_0=m k_F/\hbar^2\pi^2$, we can write this as
$n\r = n_0-g_0 v_s\r =: n_0 + n_{ind}$.
The screened potential becomes
\be
v_s\r = 
v_{ext}\r+ e^2\int d^3 r'\,\frac{n_0\rp+n_{ind}\rp}{|{\bf r}-{\bf r'}|}.
\ee
Stability of the unperturbed electron liquid (with $v_{test}=0$) requires
$n_0=-n_{bg}$, so that the background-contribution to $v_{ext}$ is 
cancelled precisely by the first term arising from the integral. Hence, 
\be
v_s\r= v_{test}\r+ e^2\int d^3 r'\,\frac{n_{ind}\rp}{|{\bf r}-{\bf r'}|}.
\ee
Upon Fourier transformation this becomes
\be
v_s({\bf k}) = v_{test}({\bf k}) + e^2{\cal F}[n_{ind}\r] 
{\cal F}[1/r],
\label{fourier3d}
\ee
where ${\cal F}$ stands for the 3-d Fourier transform operator.
In the second term we have used the convolution theorem to decompose
the Fourier transform of the Hartree potential into a product of two Fourier 
transforms. By substituting $n_{ind}=-g_0 v_s$ in this equation, using
${\cal F}[1/r]=4\pi/k^2$, and solving for $v_s$, we obtain 
\be
v_s({\bf k}) 
= \frac{v_{test}({\bf k})}{1+e^2 g_0 {\cal F}[1/r]}
= \frac{v_{test}({\bf k})}{1+{4 k_F\over a_0 \pi k^2}}
=\frac{v_{test}({\bf k})}{\epthree},
\label{3dres}
\ee
in agreement with the definitions following Eq.~(\ref{eps3def}). For the 
screening of a point charge with $v_{test}({\bf k})=4 \pi e^2/k^2$ this becomes
\be
v_s({\bf k}) = \frac{4\pi e^2 }{k^2+k_{TF}^2}.
\label{3dresPC}
\ee
Minimization of the TF kinetic-energy functional followed by a minimization 
of the resulting Euler equation thus reproduces the TF dielectric function
\be
\epthree=1+{4 k_F \over a_0 \pi} {1\over k^2}.
\label{ep3diel}
\ee
In the following sections we explore whether this connection between 
energy functionals and model dielectric functions remains valid in more
complex situations, characterized by additional exchange and correlation terms 
in the functional, and in two dimensions.

\section{TFD dielectric function in three-dimensional electron liquids}
\label{tfdsec}

The fact that dielectric
functions can be derived from energy functionals opens up the possibility
to incorporate additional physics into a model dielectric function by
using a more complete energy functional. To illustrate this we now 
deduce an expression that incorporates exchange effects into the 
TF dielectric function. Starting point is the 3-d TFD energy functional
\be
E_{TFD}[n]= E_{TF}[n] + E_x^{LDA}[n],
\ee
where the functionals on the rhs are defined in Eq.~(\ref{etotal3}) 
and (\ref{dirac3d}), respectively. The Euler equation corresponding
to this functional is
\be
\frac{5}{3}Cn\r^{2/3}+v_{ext}\r + v_H\r -\frac{4}{3}D n\r^{1/3} -\mu=0.
\ee
In accordance with common practice in density-functional 
theory\cite{dftbook,parryang} we regard the exchange term on the rhs as a 
contribution to the effective potential, now defined as
\be
v_s=v_{ext}+v_H+v_x= v_{ext}+v_H - (4D/3) n^{1/3}.
\ee
The induced density change is defined in terms of this potential as before,
via $n\r = n_0-g_0 v_s\r =: n_0 + n_{ind}\r$, and the counterpart to 
Eq.~(\ref{fourier3d}) becomes
\be
v_s({\bf k}) = v_{test}({\bf k}) + e^2{\cal F}[n_{ind}\r]
{\cal F}\left[{1\over r}\right] - {4D\over 3} {\cal F}[n^{1/3}].
\ee
To evaluate this expression in closed form we linearize again, assuming
consistently that $n_{ind} \ll n_0$, i.e., the self-consistent modification 
of the density distribution due to the test charge is small compared to the 
density of the original unperturbed system. This allows us to write
$n^{1/3} \approx n_0^{1/3}(1+n_{ind}/3 n_0)$, and the preceding equation
becomes
\bea
v_s({\bf k}) = v_{test}({\bf k}) + e^2{\cal F}[n_{ind}\r]
{\cal F}\left[{1\over r}\right] 
\nonumber \\
- {4D\over3} {\cal F}[n_0^{1/3}]
- {4D\over9n_0^{2/3}} {\cal F}[n_{ind}].
\eea
The last term can be handled as before, by using $n_{ind}=-g_0 v_{sc}$.
Since $n_0$ is spatially constant, the Fourier transform of the
second-to-last term contributes only at $k=0$.
Solving the preceding equation for $v_s$ we find
\be
v_s({\bf k}) = \frac{v_{test}({\bf k}) - 4D {\cal F}[n_0^{1/3}]/3}
{1+ \frac{k_F}{\pi^2 a_0} {\cal F}\left[{1\over r}\right] 
- \frac{1}{\pi a_0 k_F}},
\ee
where we have substituted the explicit expressions for $g_0$, $n_0$ and $D$
in terms of universal functions and the Fermi wave vector. To identify the
dielectric function we consider again the screening of a point charge,
with $v_{test}=4 \pi e^2/k^2$. After a little algebra we find
\be
\epdthree = 1+\frac{4 k_F}{\pi a_0 k^2} - \frac{1}{\pi a_0 k_F}
= \epthree - \frac{\alpha r_s}{\pi},
\label{epd3diel}
\ee
where in the last step we have introduced the usual electron gas parameters
$r_s=1/a_0 k_F \alpha$ and $\alpha=(4/9\pi)^{1/3}$. Eq.~(\ref{epd3diel})
illustrates how exchange-corrections modify the TF dielectric function, and 
also illustrates how by starting from suitable energy functionals additional 
physics can be built into models for $\epsilon({\bf k})$. 

Interestingly, the resulting additive exchange correction to the TF dielectric 
function, $-\alpha r_s/\pi$, turns out to be the same already known to appear 
as the multiplicative Hartree-Fock correction to the sound velocity ($s$), 
compressibility ($\kappa$) and spin susceptibility ($\chi$) of the 3-d 
electron gas:\cite{pinesnozieres,quantliq} 
\be
\epdthree-\epthree = 1-{s^2\over s_0^2} = 1-{\kappa_0^2\over \kappa^2}
=1-{\chi_0 \over \chi} = - \frac{\alpha r_s}{\pi}.
\label{chain1}
\ee

\begin{figure}
\centering
\includegraphics[height=65mm,width=80mm,angle=0]{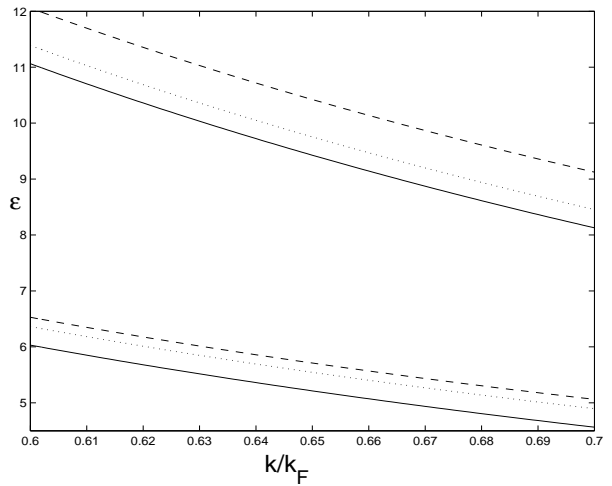}
\caption {\label{fig1} Three-dimensional Thomas-Fermi (dashed curves) 
and Thomas-Fermi-Dirac (continuous curves) dielectric function vs. wave 
vector $k/k_F$, for $r_s=6$ (upper curves) and $r_s=3$ (lower curves). 
For comparison purposes 
we also include the $k\to 0$ limit of the static RPA dielectric function
(dotted curves), as given in Eq.~(\ref{rpa}). For the density dependence
of the TF and TFD dielectric functions see Fig.~\ref{fig3}.}
\end{figure}

The behaviour of $\epdthree$ for two different densities
is illustrated in Fig.~\ref{fig1}. Exchange is seen to reduce screening (i.e., 
to lead to a dielectric function closer to $\epsilon =1$ for all ${\bf k}$).
This behaviour is easily understood because exchange tends to keep particles 
apart, thereby making screening less effective. 
The limits of validity of Eq.~(\ref{epd3diel}) 
are similar to that of ordinary TF theory, restricting the use of 
(\ref{epd3diel}) to high and slowly varying densities (small $r_s$ and
$k\to 0$).

\section{Long wavelength limit of the random-phase approximation}
\label{limit}

In the random-phase approximation the static dielectric function is 
given by\cite{ashcroft,ziman,phillips,mahan,pinesnozieres,quantliq}
\bea
\epsilon_{RPA}(k) =
1+{k_{TF}^2\over 2k^2}\left[ 
1+{1\over 2x}(1-x^2) \ln\left|\frac{1+x}{1-x} \right | \right],
\label{rpadiel}
\eea
where $x=k/2k_F$. 
In textbooks\cite{pinesnozieres,quantliq,ziman,phillips,jonesmarch,mbt} it is 
usually argued that in the limit $x\to 0$ the term in square brackets tends to 
1, because, to linear order in $x$, $\ln\left|\frac{1+x}{1-x}\right|\to2x$, 
so that the RPA dielectric function for long wavelength density 
oscillations becomes $\epsilon_{RPA}(k\to 0) = \epthree$ and thus recovers 
the TF result. (See, e.g., Eq.~(5.65) of Ref.~\onlinecite{pinesnozieres},
p. 105 of Ref.~\onlinecite{jonesmarch}, p. 140 of Ref.~\onlinecite{phillips},
Eq.~(5.35) of Ref.~\onlinecite{quantliq}, or Eq. (28.9) of 
Ref.~\onlinecite{mbt}.)

This statement, however, is not fully correct. More precisely, it is not 
consistent in orders of $k$. To the order in $k$ to which TF 
theory provides an approximation to $\epsilon({\bf k})$, i.e., constant plus
$1/k^2$, it is not enough to expand the logarithm in Eq.~(\ref{rpadiel}) 
to linear order in $x$. Instead, the expansion must be carried 
to cubic order, which after multiplication by the prefactor 
${1\over 2x}(1-x^2)$ also contributes a constant term to $\epsilon^{RPA}(k)$. 
The RPA dielectric function thus has the $k\to 0$ expansion
\bea
\epsilon_{RPA}({\bf k}\to 0)
={4 k_F \over a_0 \pi} k^{-2} 
+ \left(1-{1\over 3} \frac{\alpha r_s}{\pi}\right)k^0 
+ O(k^2)
\\
=\epthree
-{1\over 3} \frac{\alpha r_s}{\pi} + O(k^2).
\label{rpa}
\eea
In the extreme $k\to0$ limit all constants can be neglected compared to
$1/k^2$, and RPA and TF theory do indeed predict the same divergence.
However, already at the level of constant corrections to this limit, RPA 
and TF differ, and the TF dielectric function in the form
$1+{k_{TF}^2/k^2}$ is never obtained as a long wavelength limit from
the RPA (except in the extreme case $r_s=0$). 

Interestingly, the additional term $-\alpha r_s/3\pi$ needed to make the 
$k\to 0$ expansion of the static RPA consistent in orders of $k$ is, up to 
a factor $1/3$ the same additional term we found above as an exchange 
correction to the TF dielectric function.


\section{Correlation corrections to the Thomas-Fermi dielectric function}
\label{tfdcsec}

Further many-body effects can be included by adding the local-density
approximation for correlation, Eq.~(\ref{corrlda}) to the TFD energy 
functional. We introduce the derivatives
\be
v_{xc}[n]\r=\frac{\delta E_{xc}[n]}{\delta n\r}
\ee
and
\be
f_{xc}[n]({\bf r},{\bf r'})=\frac{\delta^2 E_{xc}[n]}{\delta n\r \delta n\rp}.
\ee
and note that in the LDA, Eq.~(\ref{corrlda}), the exchange-correlation kernel 
$f_{xc}$ is
\bea
f_{xc}^{LDA}[n]({\bf r},{\bf r'}) = 
\frac{\partial^2e_{xc}(n)}{\partial n^2}|_{n\to n\r} \delta({\bf r}-{\bf r'})
\\
=:\bar{f}_{xc}^{LDA}[n]\delta({\bf r}-{\bf r'}),
\eea
where the second equality defines the functional $\bar{f}_{xc}^{LDA}[n]$.
The Fourier transform of $f_{xc}^{LDA}$ is simply related to that of the
static local-field factor ${\cal G}({\bf k})$ via \cite{dftbook}
\be
f_{xc}^{LDA}[n]({\bf k})
= - e^2{\cal F}\left[{1\over r}\right] {\cal G}({\bf k}),
\ee
where $-e^2{\cal F}[1/r]$ is the Fourier transfrom of the Coulomb interaction.

By going through the same steps as before we find for the dielectric
function corresponding to correlation-corrected TFD theory
\bea
\epcthree = 1+\frac{4 k_F}{\pi a_0 k^2} + g_0 \bar{f}_{xc}^{LDA}[n_0]
\\
= \epthree + g_0 \bar{f}_{xc}^{LDA}[n_0].
\label{epc3diel}
\eea
Interestingly, the inclusion of correlation, already on the linearized LDA 
level, has introduced the exchange-correlation kernel $f_{xc}$ in the 
corrected TF dielectric function, and thus lead to a dependence on local-field 
corrections that would ordinarily only be expected in sophisticated
beyond-RPA dielectric functions.\cite{pinesnozieres,quantliq,mahan} 
The wave-vector dependence of the local-field correction, however, is
lost in the linearization.

\section{TF dielectric function in two-dimensional electron liquids}
\label{2dtfsec}

\begin{figure}
\centering
\includegraphics[height=65mm,width=80mm,angle=0]{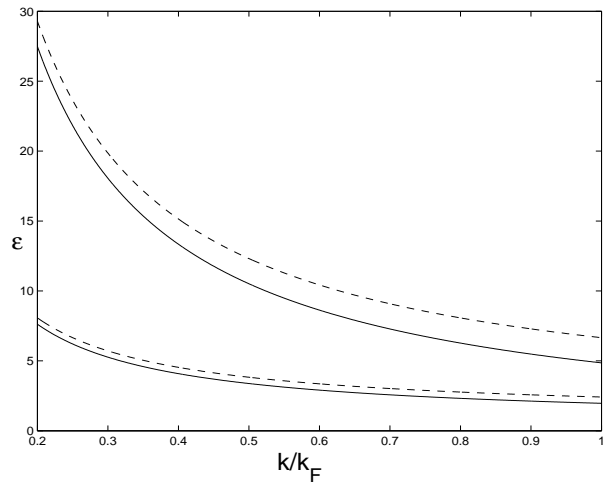}
\caption {\label{fig2}Two-dimensional Thomas-Fermi (dashed curves) and
Thomas-Fermi-Dirac (continuous curves) dielectric function vs. wave vector
$k/k_F$, for $r_s=4$ (upper curves) and $r_s=1$ (lower curves). Inclusion
of exchange shifts the dielectric function to lower values, the shift being
larger for lower densities (larger $r_s$), and for fixed density larger in
two than in three dimensions. For the density dependence of the 2-d TF and 
TFD dielectric functions see Fig.~\ref{fig3}.}
\end{figure}

Next, we extend the above analysis to two dimensions, where
our starting point is the 2-d local-density approximation to the
kinetic-energy functional,
\be
\ttftwo=\frac{\pi \hbar^2}{2m} \int d^2 r \, n\r^2.
\ee
(Here and below 2-d energy functionals and dielectric functions 
are distinguished from their 3-d counterparts by a subscript $2$. For
all other quantities no confusion can arise, and to avoid cluttering the 
formulas we use the same symbol for them in 2-d as in 3-d.)
By following exactly the same steps as before we find the Euler equation
\be
\frac{\pi \hbar^2}{m} n\r + v_{ext}\r + v_H\r - \mu =0.
\ee
Unlike Eq.~(\ref{eulern}) in 3-d, this density-potential relation is already
linear, and no further approximation is needed to write it in the form
$n\r=n_0+n_{ind}\r$. The 2-d Fermi wave vector is related to the 2-d density
via $n_0=k_F^2/2\pi$, and in terms of the 2-d density of states 
$g_0=m/\pi\hbar^2$ one immediately identifies $n_{ind}=-g_0 v_s$. All 
remaining steps are the same as in 3-d, and from the 2-d Fourier transform 
${\cal F}[1/r]=2\pi /k$ one readily obtains
\bea
v_s({\bf k}) =
\nonumber \\
\frac{v_{test}({\bf k})}{1+e^2 g_0 {\cal F}[1/r]}
=\frac{v_{test}({\bf k})}{1+{2\over a_0}{1\over k}}
=\frac{v_{test}({\bf k})}{\eptwo}.
\label{2dres}
\eea
For screening of a point charge (\ref{2dres}) becomes
\be
v_s({\bf k}) = \frac{2\pi e^2 }{k+k_{TF}},
\label{2dresPC}
\ee
where in 2-d $k_{TF}=2/a_0$.
The resulting density-independent expression
\be
\eptwo=1+{2\over a_0}{1\over k}
\label{ep2diel}
\ee
for the 2-d TF dielectric function is well known, e.g., in the context of
2-d electron liquids in semiconductor heterostructures.\cite{stern1,andormp}

Interestingly, the 2-d TF dielectric function follows rigorously from
the 2-d TF kinetic energy functional, without an additional linearization.
Hence, in this sense, (\ref{2dres}) is a stronger result than its 3-d 
counterpart (\ref{3dres}).\cite{footnote1} However, this stronger result is 
an artifact of two-dimensionality, arising because the 2-d kinetic energy 
is quadratic in $n\r$, and its density-derivative linear.

On the other hand, the fact that $\eptwo$ is obtained from $\ttftwo$ in the 
same way $\epthree$ is obtained from $\ttfthree$ shows that the relation 
between energy functionals and dielectric functions in itself is not an 
artifact of three-dimensionality.\cite{footnote2}

\section{TFD dielectric function in two-dimensional electron liquids}
\label{2dtfdsec}

Finally, we work out the form of the exchange corrections to the
two-dimensional Thomas-Fermi dielectric function.
The per-particle exchange energy of a uniform two-dimensional 
electron liquid is $-8 \sqrt{2}/(3 \pi r_s) Ry$.\cite{stern2,isihara} where
$r_s=1/(a_0\sqrt{ n \pi})$ and $Ry=e^2/2a_0$. The local-density approximation
for $E_{x,2}[n]$ is thus
\be
E_{x,2}^{LDA}[n]=-{4 \over 3} \sqrt{2 \over \pi} e^2 \int d^2 r\, n^{3/2}.
\ee
The 2-d TFD Euler equation can be cast in the same form as before,
$n\r = n_0 + n_{ind}\r$, where in 2-d $n_0=k_F^2/2\pi$ and 
$n_{ind}=-g_{0,2} v_s$, with $v_s=v_{ext}+v_H+v_x$ and 
$v_x=-2 e^2 \sqrt{2 n/\pi}$. The Hartree 
potential arising from the unperturbed density $n_0$ and the potential energy
in the positive background (contained in $v_{ext}$) cancel again, and by means
of the convolution theorem the Fourier-transformed screened potential becomes
\be
v_s({\bf k}) = v_{ext}({\bf k}) 
+ e^2{\cal F}_2[n_{ind}\r] {\cal F}_2\left[{1 \over r}\right]
-2 e^2 \sqrt{2\over \pi} {\cal F}_2[\sqrt{n}].
\ee
The Fourier transforms in the second term on the rhs are easily evaluated by 
using $n_{ind}= -g_{0,2} v_s({\bf k})$ and ${\cal F}_2[1/r]=2\pi/k$. To 
evaluate the Fourier transforms in the third term we linearize in terms of 
$n_{ind}/n_0$, as before, and obtain
\be
{\cal F}_2[\sqrt{n_0+n_{ind}}] \approx {\cal F}_2[\sqrt{n_0}] 
- {1\over 2 \sqrt{n_0}} {\cal F}_2[n_{ind}].
\ee
Hence, for the 2-d screening of a point charge with 
$v_{test}({\bf r})=e^2/r$, we obtain
\be
v_s({\bf k}) = \frac{2 \pi e^2}{k}
\frac{1}{1+{2\over a_0}{1\over k} -{\sqrt{2}\over \pi} r_s},
\ee
where we have used $r_s =\sqrt{2}/a_0 k_F$ for the 2-d
density parameter $r_s$, and, as before, $k\,\delta(k)\equiv 0$. 
As in the 3-d case, the derivation from a kinetic-energy functional allows 
us to build exchange effects in a simple way into the TF dielectric function. 

\begin{figure}
\centering
\includegraphics[height=65mm,width=80mm,angle=0]{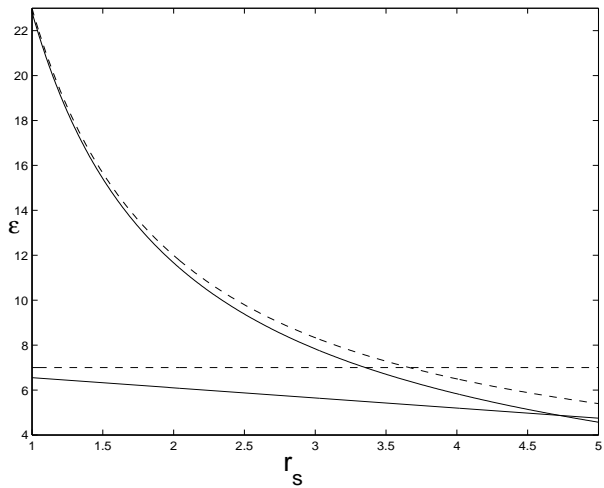}
\caption {\label{fig3} Density dependence of the 3-d TF (dashed line),
3-d TFD (full line), 2-d TF (dotted line) and 2-d TFD (dash-dotted line)
dielectric functions at fixed wave vector $k=1/3a_0$ (where
$\lambda =2\pi/k=6 \pi a_0$ is the wave length of the density variations
in $n_{ind}\r$). The change of the TF dielectric function due to exchange
corrections is a downshift that is more pronounced at low densities (large
$r_s$). In 2-d it is only because of the exchange corrections that a
density-dependence appears in the dielectric function.}
\end{figure}

Interestingly, our final expression 
\be
\epdtwo = 1+{2\over a_0}{1 \over k} -{\sqrt{2}\over \pi}r_s
=\eptwo  -{\sqrt{2}\over \pi}r_s
\label{epd2diel}
\ee
has aquired a density dependence, through $r_s(n)=1/a_0\sqrt{ n \pi}$. 
Inclusion of exchange thus removes the unphysical density-independence of 
$\eptwo$ obtained in the 2-d TF approximation (\ref{2dres}). 
The behaviour of the 2-d TF and TFD dielectric functions as a function 
of wave vector and density is illustrated in Figs.~\ref{fig2} and \ref{fig3}.
Correlation corrections could be included in two dimensions in the same way 
as in three dimensions. In the interest of brevity we refrain from showing 
the straightforward extension of the above expressions to this case.

On the other hand, we stress that in 2-d we obtain the same chain of
equalities as in 3-d,  relating the additive corrrection to the dielectric 
function with multiplicative renormalization of the sound velocity, 
compressibility and spin susceptibility of the 2-d electron 
gas:\cite{polini1,polini2} 
\be
\epdtwo-\eptwo= 1-{s^2\over s_0^2} = 1-{\kappa_0^2\over \kappa^2}
=1-{\chi_0 \over \chi} = -{\sqrt{2}r_s\over \pi}.
\label{chain2}
\ee
Eq.~(\ref{chain1}) thus carries over to 2-d, with a modified density 
dependence.

\section{Summary and assessment}
\label{summary}

While the 2d and 3d TF dielectric functions $\epthree$ and $\eptwo$, given in 
Eqs.~(\ref{ep3diel}) and (\ref{ep2diel}), are well known (although not
normally derived from density functionals), their TFD 
counterparts $\epdthree$ and $\epdtwo$, Eqs.~(\ref{epd3diel}) 
and (\ref{epd2diel}), and the correlation-corrected expression
(\ref{epc3diel}) are apparently derived for the first time in this work.
We do not expect simple models of this type to substitute
the sophisticated models available from diagramatic many-body 
physics,\cite{mahan} or the material-specific dielectric functions that can be
obtained from density-functional theory.\cite{abineps} 
However, each of the four extensions of the 3-d TF functional studied here 
displays interesting features that were not obvious from the 3-d TF case, 
but are revealed by deriving model dielectric constants from density 
functionals.

Specifically in the 3-d TFD case, we found that the exchange contribution 
to the TF dielectric function yields a term of the same form as the 
long-wavelength expansion of the RPA, if that expansion is carried out 
consistently in orders of $k$. The standard way of identifying $\epthree$ 
as the $k\to0$ limit of the static RPA is not consistent in orders of $k$. 
Further correlation terms lead, already on the linearized LDA level, to a 
simple dependence of the dielectric function on local-field factors. 

Unlike the connection between the TF dielectric function and the Yukawa
potential in real space, that between the TF dielectric function and
energy functionals carries over to two-dimensions, where $\eptwo$ turns out
to be a more robust expression than $\epthree$ in 3-d, because it can be 
derived from the 2-d TF kinetic-energy functional without any linearization. 
Upon including exchange (TFD), a density dependence reappears in the 2-d 
dielectric function, which is lost in the usual 2-d TF dielectric function.
In both 2d and 3d, the additive exchange contribution to the dielectric 
function turns out to be the same as the multiplicative exchange 
renormalization factor of other observables, such as the spin susceptibility.

The existence of a connection between a class of energy functionals 
and a set of model dielectric functions opens the question whether more 
detailed models for $\epsilon({\bf k})$ can be obtained from the more 
complex energy functionals in use in modern density-functional theory.
Models for frequency-dependent dielectric functions may be accessible via
time-dependent density-functional theory.\cite{tddft}

{\bf Acknowledgments}\\
This work was sup\-por\-ted by FAPESP, CNPq and FUNDECT. We thank 
Lid\'erio C. Ioriatti and Marco Polini for useful discussions.

\end{document}